**Applications of Enhanced Sampling Methods to Biomolecular Self-Assembly: A Review**

*Mason Hooten\*,1, Het Patel\*,2, Yiwei Shao\*,2, Rishabh Kumar Singh\*,2, and Meenakshi Dutt\*\*,2*

[1] Biomedical Engineering, Rutgers, The State University of New Jersey, Piscataway, New Jersey 08854
[2] Chemical and Biochemical Engineering, Rutgers, The State University of New Jersey, Piscataway, New Jersey 08854
\* equal contributors
\*\* corresponding author: meenakshi.dutt@rutgers.edu

**Abstract**

This review article discusses some common enhanced sampling methods in relation to the process of self-assembly of biomolecules. An introduction to self-assembly and its challenges is covered followed by a brief overview of the methods and analysis for replica-exchange molecular dynamics, umbrella sampling, metadynamics, and machine learning based techniques. Applications of select methods towards peptides, proteins, polymers, nucleic acids, and supramolecules are discussed. Finally, a short discussion of the future directions of some of these methods is provided.

# Section 1 - Introduction

The self-assembly of biomolecules to form biomolecular assemblies and materials is ubiquitous both in nature and the laboratory. A mechanistic understanding of molecular self-assembly is essential to a wide range of disciplines within biomedical sciences for applications such as crystallization or drug delivery (Whitesides and Grzybowski 2002). The molecular mechanisms underlying the process of self-assembly can be elucidated via molecular simulation techniques such as Molecular Dynamics (MD). The spatiotemporal scales resolved by these simulation techniques with current computational resources are limited when attempting to explore processes with extended time scales (Hollingsworth and Dror 2018). These limitations can lead to insufficient sampling of a broad range of conformations (Bernardi et. al. 2015) and configurational states, thereby impacting characterization of the system. To overcome these challenges, enhanced sampling methods have been employed to provide a deeper understanding of the mechanisms and pathways underlying the self-assembly process. This is done by parsing the jagged free energy landscape of biomolecules and properly sampling a broad range of conformations and configurations (Bernardi et. al. 2015). This review covers a selection of enhanced sampling techniques that have been used to provide a mechanistic insight into the process of self-assembly of biomolecules.

# Section 2: Background and Motivation

**Self-Assembly: Definition**

Self-assembly is the autonomous organization of components into structured, functional arrangements without external guidance (Whitesides and Grzybowski 2002). Self-assembly falls into two primary types: static and dynamic. In static self-assembly, structures remain stable once formed, driven by minimization of the free energy. Conversely, dynamic self-assembly requires continuous energy input, allowing structures to exist transiently or evolve over time. Yet, while thermodynamics often drives the process by favoring lower free energy states, the resulting structures may not always represent global or even local minima of the free energy.

Whitelam's (Whitelam and Jack 2014) definition of self-assembly focused on it being an intrinsically disordered nonequilibrium process. Near-equilibrium assembly emphasizes thermodynamic factors, whereas far-from-equilibrium assembly relies heavily on dynamic effects, where conditions such as controlled parameters are crucial to achieving stable ordered structures. Computational approaches are essential for studying self-assembly due to their ability to resolve the underlying processes and mechanisms arising from the interplay between key physical interactions. Advanced algorithms and rare-event sampling techniques, such as forward-flux and transition-path sampling, enhance the sampling of complex systems, especially those with multiple ordered structures and vital for capturing transient phenomena. Despite their inherent approximations, such as coarse-grained models and simplified solvent interactions, computational approaches provide valuable qualitative insights that complement experimental findings, such as those associated with hydrophobic or hydrodynamic effects (Ardekani et al. 2024). Hence, these approaches provide an efficient and low-cost avenue for designing and predicting the formation and behavior of complex systems, which may be expensive or challenging to replicate experimentally.

These self-assembly processes, governed by molecular and supramolecular interactions, are central to various natural and synthetic systems, each characterized by unique thermodynamic and kinetic principles:

(1) Self-assembly of peptides involves the spontaneous organization of amino acid sequences into ordered nanostructures, such as fibrils, micelles, and hydrogels. Thermodynamic driving forces include enthalpic contributions from hydrogen bonding and van der Waals interactions, as well as entropic gains from solvent structuring and chain mobility. (Tantakitti 2016; Freeman 2018) Yan et al. (2016) noted that such a process is primarily driven by thermodynamics, characterized by various non-covalent interactions, and is significantly modulated by kinetics, due to external conditions such as salt concentration, pH and temperature. The kinetics of peptide self-assembly are influenced by nucleation mechanisms and growth rates, which determine the pathway and morphology of the resultant structures (Zhang et al. 2019; Chen et al. 2022). The interplay of thermodynamics and kinetics often contribute to kinetic adjustments that disrupt the equilibrium of the thermodynamic driving forces or alter energy barriers, ultimately leading to a kinetically trapped state.
(2) In polymeric systems, self-assembly enables the formation of nanostructured materials with tailored properties. Yan et al. (2013) noted that hierarchical arrangements due to intermolecular interactions define the properties of polymer nanocomposites, while the movement of nanoparticles affects the behavior of the polymer components. Key factors in these systems include: the interaction between nanoparticles and polymers, long-range forces (such as van der Waals or electrostatic forces) between particles, nanoparticle size and shape, and the composition and molecular structure of the polymers. The challenge in designing these materials lies in understanding how to leverage the intricate balance between entropy and enthalpy to achieve the resulting structures. Computational approaches have provided insights into the thermodynamic stability and kinetic pathways underlying self-assembly, particularly under external fields such as electric or magnetic fields. Interactions between polymer chains and nanoparticles have contributed to the formation of hierarchical structures (Yadav et al. 2020; Bhendale and Singh 2023).

(3) Self-assembly of lipids results in the formation of micelles, bilayers, and other structures, governed by the hydrophobic effect and molecular packing parameters. These systems are known to be highly sensitive to external stimuli, including temperature, pH, and ionic strength (Zhang et al. 2014; Grzelczak et al 2019). One prominent subarea of study, colloidal self-assembly, leverages inter-particle dynamics, including but not limited to van der Waals interactions, electrostatic interactions, hydrophobic forces and solvation force. Li et al. (2008) suggested that the many-body interactions of coupled particles are better described in their molecular models compared to continuum theories, which are normally accompanied by adequate approximation to reduce the computational cost of simulations. Similarly, crystallization-driven self-assembly, often mediated by block copolymers, enables the formation of well-defined structures with precise functional properties (Nguyen et al. 2021). Polymer-assisted crystallization and crystallization-driven self-assembly are influenced by the thermodynamics and kinetics of biomolecular interactions, including dimerization. Zheng et al. (2021) investigated the interplay of factors like entropy, enthalpy, and molecular architecture as crucial parameters for controlling the assembly process.

(4) Self-assembly of proteins underpins numerous biological phenomena, including amyloid fibril formation and viral capsid assembly. The thermodynamics of protein assembly involve the minimization of free energy through hydrophobic and electrostatic interactions. Kinetic factors, such as nucleation rates and conformational changes, determine the efficiency and specificity of assembly processes. (Zerovnik et al. 2016). For example, Asherie et al. (2016) reported that the formation of amyloid fibrils involved crucial conformational changes and various pathways. Tezcan et al. (2021) highlighted that protein self-assembly, distinct from protein folding, can be pathway-dependent, resulting in varying structural outcomes depending on the environmental conditions. While computational modeling is a powerful tool in the design of supramolecular protein aggregates, it often relies on a trial-and-error approach to address challenges related to polar interactions and dynamic pathways.

**Self-assembly: Underlying Processes**

The self-organization of molecules into assemblies naturally follows through complex hierarchical pathways. Beyond the prescription of building patterned structures from their constituents, one by one, to a certain template, most self-assemblies are formed through the aggregation of clusters, with multiple non-covalent intermolecular interactions – like hydrogen bonding, pi-pi stacking, electrostatics and van der Waals forces – underlying the process. Self-assembly, driven without any external energy supply, is mainly guided by thermal fluctuations from the solvent molecules. The components fluctuate and undergo conformational changes through diffusion in solution without dissipating energy (Whitelam and Jack 2015).

Dispersed monomers in the solution adhering to statistical mechanics are ideally expected to achieve conformations with ever-lower free energy and form assemblies but this is rarely witnessed on experimental timescales. Rather some components first self-assemble as metastable structures that may or may not relax to form a thermodynamically stable assembly on experimental or computational timescales. As discussed earlier, self-assembly can happen either near equilibrium or far from equilibrium. While near-equilibrium self-assembly is thermodynamically controlled, far-from-equilibrium self-assembly is kinetically controlled. Most static assemblies that require no dissipation of energy are near equilibrium. Whereas dynamic assemblies dissipate energy, going downhill far from equilibrium in the free energy landscape (Whitesides and Grzybowski 2002). Though far from equilibrium self-assembly is kinetically frustrated, the interplay of stronger interaction driven by large fluctuation and specificity for productive binding leads to faster kinetics than near equilibrium. Whereas reversibility or detailed balance and low energy impetus slows down the kinetics.

As revealed by experiments and computer simulations, two important factors act as control parameters for many self-assemblies: 1) thermodynamic favorability towards forming an ordered, stable structure, and 2) conditions allowing the random arrangements to organize into an ordered structure. Such requirements may be inhibited due to an imbalance in the strength of interaction and specificity of binding. If the interactions between components are too specific, it impedes productive binding. Non-specific interactions lead to partial reversibility whereby intermediate structures can transiently break bonds to correct the nascent defects of growing assemblies. This error correction mechanism makes for a necessary condition for robust self-assembly (Whitelam, Feng et al. 2009; Whitelam 2010). Overly strong interactions do not allow for adequate relaxation time leading to malformed aggregates. The non-specific interactions lead to kinetic frustration while the specific interaction leads to thermodynamic frustration, hence the competition between thermodynamics and kinetics decides the viable phase space for self-assembly (Whitelam, Feng et al. 2009; Whitelam 2010).

The timescales to explore self-assembly pathways in molecular simulations are usually limited to the nanosecond range with a system size of a few 100 monomers (Frederix, Patmanidis, et al. 2018). Most all-atom simulations are unable to capture all the dynamic stages in a self-assembly process. Enhanced sampling techniques help capture not only the overall morphologies of thousands of monomers but also the various manifolds of kinetic pathways from one structure to the other. The main idea is to use biasing potentials along certain reaction coordinates (RCs) to effectively reduce the energy barrier on the free energy landscape such that all relevant thermodynamics states are spanned. The choice of collective variables (CVs) and the enhanced sampling method are of immense importance in reconciling the complex interplay of thermodynamics and kinetics.

# Section 3 - Methods and Theory

**Replica-Exchange Molecular Dynamics**
Replica exchange (RE) is an algorithm for enhancing sampling by exchanging structural information between simulations, in an ensemble of simulations that vary in terms of an instrumental parameter. This parameter usually modulates system energy, with the effect that RE is able to overcome high energetic barriers, sampling broadly over energy landscapes with many degrees of freedom where traditional sampling may encounter kinetically trapped microstates.

The application of RE to MD (REMD) (Sugita 1999; Nymeyer 2004) is a common strategy for investigating biomolecules. In REMD, a set of MD parameterizations representing an ensemble of thermodynamic states is prepared in which each state parameterization takes on a different value of a particular parameter, often temperature. Chemically identical systems of atoms – termed replicas – are simulated in parallel in the ensemble of states, and periodically attempts are made to exchange system configurations between pairs of states.

For example, consider a pair of thermodynamic states $m$ and $n$ with reciprocal temperature $\beta_m$ and $\beta_n$, respectively. Prior to an exchange, these states contain atomistic configurations $x$ and $y$, respectively. The probability of exchanging $x$ and $y$ between $m$ and $n$ is calculated as a function of the potential energy $E$, such that swaps will only be completed if the end state is consistent with the target thermodynamic ensemble. Equation (1) expresses the weighted difference in potential energy between replicas, and an exchange attempt is accepted or rejected via the Metropolis criterion shown in equation (2).

$$\Delta \equiv [\beta_n - \beta_m](E(x) - E(y)) \tag{1}$$

$$P_{exchange}(x, y) \equiv \begin{cases} 1 & for\ \Delta \leq 0 \\ exp(-\Delta) & for\ \Delta > 0 \end{cases} \tag{2}$$

In the typical REMD simulation, repeated rounds of MD are punctuated by exchange steps, shown schematically in Figure 1. The result of these exchange steps is that each system of atoms conducts a random walk through the prescribed temperature space, as exemplified in Figure 2. The result is a more robust exploration of the energy surface, and therefore the structural ensemble, at lower energy states.

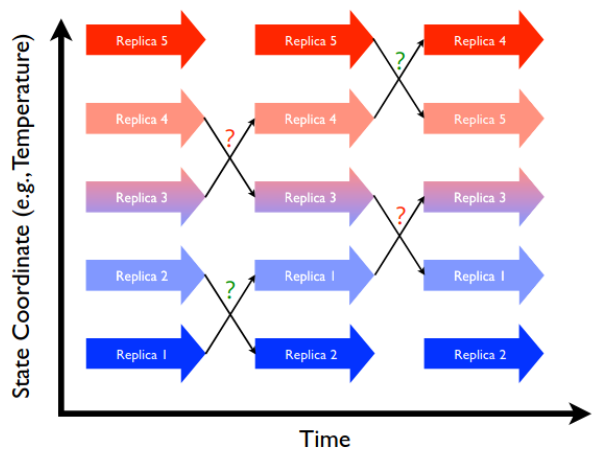

Figure 1. Schematic representation of REMD showing MD time elapsing between periodic RE attempts. Reprinted from the Amber reference manual (Case 2005; Amber Manuals 2025).

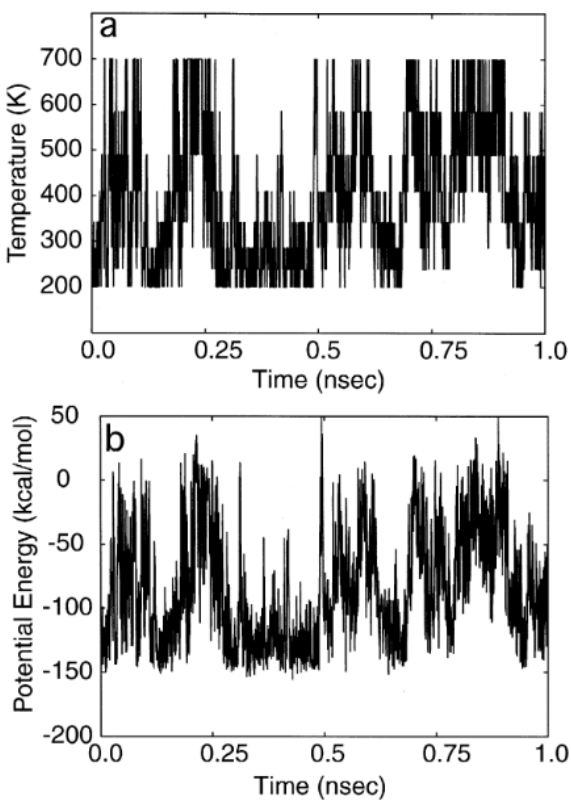

Figure 2. Time series of (a) temperature and (b) potential energy for a single replica over the course of a REMD simulation. Reprinted from reference (Sugita 1999) with permission from Elsevier.

Temperature is the most common dimension over which replicas are designed to vary, but REMD may be implemented using a variety of other dimensions including charge states, center of mass (COM) distance in umbrella sampling, and the simulated Hamiltonian. REMD of unconstrained

MD simulations has led to significant results in the study of self-assembly of biomolecules such as peptides, lipids, and proteins. The combination of REMD with other techniques, including umbrella sampling, has also been used to characterize the energy of interaction between individual molecules in targeted structures of interest such as amyloid protofibrils and peptides embedded in lipid membranes. Although REMD is the most commonly reported RE approach in biomolecular simulations, there are examples of RE applied to a variety of stochastic and dynamical simulation techniques, including Wang-Landau energy sampling (Vogel 2014; Wang 2001) and dissipative particle dynamics (DPD) (Kobayashi 2019).

Temperature is by far the most commonly reported REMD parameter in studies concerning biomolecular aggregation (Anand 2008; Baumketner 2005; Cecchini 2004; De Simone 2010; Mu 2012; Rissanou 2013; Sieradzan 2012; Soto 2006; Takeda 2008; Tamamis 2009; Xiong 2019), and temperature REMD (T-REMD) is implemented in a variety of MD software packages (Abraham 2015; Brooks 2009; Case 2005; The PLUMED Consortium 2019). REMD has also been combined with umbrella sampling (see next section) in studies of oligomerization, where replicas vary on the constrained intermolecular COM distance (i.e., umbrella distance) (Martel 2017; Wolf 2008).

Hamiltonian REMD (H-REMD), extends T-REMD by directly acting on the functional form of the Hamiltonian, such that the effects of temperature increase may be confined only to a subsystem. For example, it may be desirable to enhance the structure of the solute (e.g. a protein) but not the solvent, a technique sometimes called solute tempering (ST) (Lockhart 2023). Multidimensional REMD algorithms have also been developed, where replicas vary simultaneously over multiple parameters, such as temperature plus umbrella distance (Gee 2011; Jeon 2013; Jeon 2014) or temperature plus protonation state of amino acid side chains (Morrow 2012). Other models which have been used to study biomolecular assembly with RE include dissipative particle dynamics (DPD) (Kobayashi 2004) and Monte Carlo sampling routines (Urano 2015; Vogel 2014).

**Umbrella Sampling**
An understanding of the interactions, organization and packing of the molecules within equilibrium self-assembled structures is imperative. For instance, investigations of the supramolecular self-assembly of biomolecules provide a fundamental understanding of the origins of the properties of these structures (Levin et. al. 2020). Further, it allows one to probe the effect of altering environmental conditions on the design and creation of new supramolecules which yield novel self-assembled structures with desired set of characteristics. Unfortunately, in biomolecular self-assembly it is challenging to determine which specific physical forces or chemical effects result in the formation of a given structure (Yu and Schatz 2013). This requires the estimation of the free energy as it is highly correlated to common biochemical events and interactions (Kastner 2011). Yu and Schatz stated that parsing the free energy differences of a biomolecule is crucial when attempting to study the conformational changes of molecules within self-assembled structures (Yu and Schatz 2013). Calculations of the difference in free energy of unlikely conformations, metastable states, and other rare events can be performed via umbrella sampling (Kastner 2011). Umbrella sampling was originally developed by Torrie and Valleau in 1976 as an effective means to calculate the free energy differences during phase transitions as traditional numerical integration methods were inefficient during these events (Torrie and Valleau 1976). Currently, umbrella sampling is often used to probe a variety of processes including self-assembly, with many methods and variants proposed over the years (Hansen and Gunsteren 2014). It is often a complementary technique applied alongside REMD and Metadynamics. Umbrella sampling attempts to sample unlikely molecular conformations and overcome steep potential wells by leveling the energy landscape through the use of a biasing potential. This bias is an additional term in the potential energy equation that should ideally make it easier for a molecule to explore various regions of the potential energy landscape in the duration of a MD simulation (Mills and Andricioaei 2008). Unfortunately, the potential energy landscape is unknown beforehand, so determining the bias that should be added is challenging. Harmonic biases are frequently used, however there are other variants including an adaptive bias (Kastner 2011). Finally, the sampling technique is applied on a RC of choice. The sampling method can be performed on the entire range of the RC or the RC can be split into increments and a simulation can be run on each increment. These split increments and independent simulations are known as windows which can contain their own bias and are eventually combined and averaged using histogram techniques such as the weighted histogram analysis method (WHAM). Taken from Justin A. Lemkul's GROMACS tutorials website, Figure 3 shows a nice illustration of the concept of windows.

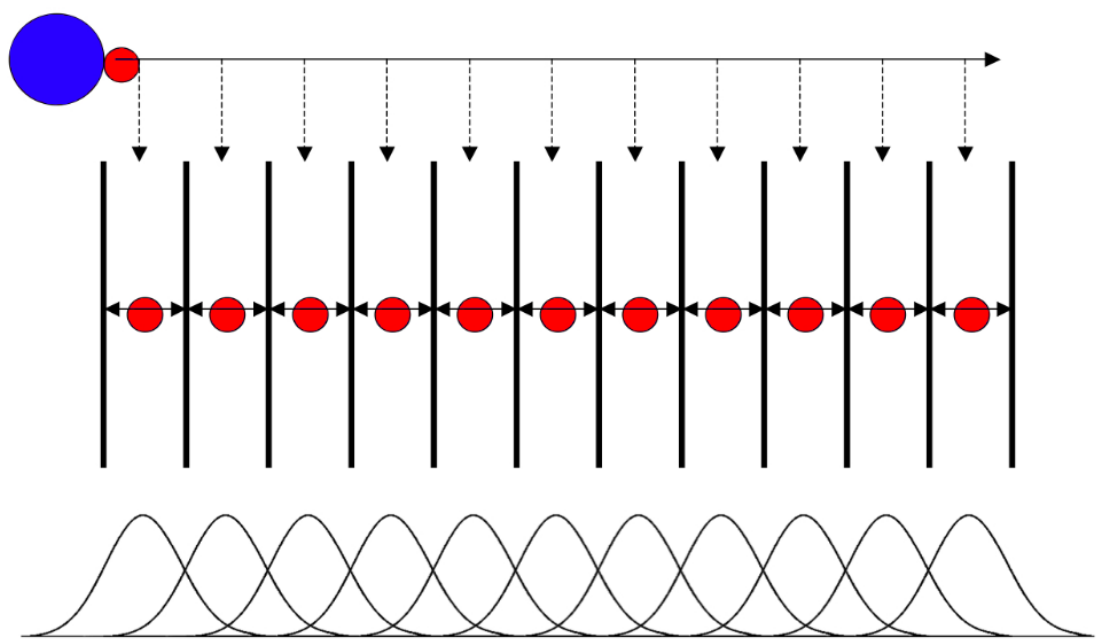

Figure 3. Umbrella sampling schematic of molecule pulling (Lemkul 2018).

In Figure 3, the red sphere is being slowly pulled away from the blue sphere. The reaction coordinate is the distance between spheres and the entire distance range to be studied is split into increments or "windows". In each of these windows a bias is applied and a simulation is performed. On the bottom is the ideal configuration histogram for each window, where each configuration overlaps so that a continuous energy function can be calculated from WHAM.

The concept of umbrella sampling is discussed in detail in the original paper by Torrie and Valleau. The mathematics shown here is a brief re-statement of the work presented in the review article by Kastner. The umbrella sampling bias term, W, is added towards the potential energy and depends on the RC of choice $\xi$.

$$Ebiased(r) = Eunbiased(r) + W(\xi) \qquad (3)$$

In equation (3), *r* is radius and *E* represents potential energy. Since the energy of a molecule is correlated to the Boltzman distribution, the unbiased and biased energies can be expressed in terms of probability distribution functions.

$$Punbiased = \int exp(-\beta Eunbiased)\delta(\xi(r) - r)\, d^N r \Big/ \int exp(-\beta Eunbiased)\, d^N r \qquad (4)$$

$$Pbiased = \int exp(-\beta Ebiased)\delta(\xi(r) - r)\, d^N r \Big/ \int exp(-\beta Ebiased)\, d^N r \qquad (5)$$

In equations (4) and (5), *P* denotes a probability distribution, *N* is the number of atoms in the system, and $\beta$ is the reciprocal temperature. Utilizing equation (3), the Ebiased term in equation (5) can be represented as the sum of the unbiased energy and the bias term *W*. Also, since the integral is of radius, the biased term can be taken outside the integral. This results in equation (6).

$$Pbiased = exp(-\beta W) \int exp(-\beta Eunbiased)\delta(\xi(r) - r)\, d^N r \Big/ \int exp(-\beta Ebiased)\, d^N r \qquad (6)$$

There is now a common integral term of exp(-$\beta$*Eunbiased)$\delta(\xi(r) - r)$ in equations (4) and (6), which can be used to relate the two distributions, resulting in

$$Punbiased = Pbiased$$
$$* exp(\beta W) \int exp(-\beta Eunbiased) exp(-\beta w)\, d^N r$$
$$\Big/ \int exp(-\beta Eunbiased)\, d^N r \qquad (7)$$

From statistical mechanics an average for a general property X is given by

$$<X> = \int exp(-\beta E)X\, d^N r \Big/ \int exp(-\beta E)\, d^N r \qquad (8)$$

This definition in the unbiased distribution, equation 7, is given by

$$Punbiased = Pbiased * exp(\beta W) <exp(-\beta W)> \qquad (9)$$

The Helmholtz free energy, A, is given by

$$A = -1/\beta * \ln(P) \tag{10}$$

Using the definition of the Helmholtz free energy on equation (9) followed by some algebraic manipulation results in

$$A = -1/\beta * \ln(Pbiased) - W - 1/\beta * \ln(<\exp(-\beta W)>) \tag{11}$$

Hence, the free energy can be determined from the selection of bias W added to the energy.

The difficulty in umbrella sampling lies in the selection of a bias and the RCs (Kastner 2011). Ideally the biased term would level the energy curve to allow for easier access to unlikely conformations, however, this is unknown since the goal is to calculate the free energy differences. Kastner notes that there are two commonly applied types of biasing potentials: harmonic bias and adaptive bias. Harmonic bias applies a harmonic term with tunable strength and reference displacement. The equation for this is shown below.

$$W(\xi) = 0.5K(\xi - \xi reference)^2 \tag{12}$$

It is common to adjust the constant K to a value that just allows the dynamics to overcome barriers. A very large constant will overrepresent high energy states and also result in a narrow harmonic function term with little overlap within the windows. This makes analysis and integration difficult. In harmonic bias umbrella sampling it is also common to split the range of the RC into separate windows so that the harmonic bias can be applied in each window. WHAM is commonly used both for the analysis of trajectories and calculation of free energies. Since the bias term to be added is unknown, another approach is to begin with an initial guess and iteratively adjust the bias so that it leads to a flatter energy landscape. This is commonly known as adaptive bias.

**Metadynamics**
While umbrella sampling effectively uses a fixed harmonic bias potential W to focus sampling on specific regions of the configuration space for systems with known energy barriers, it shows limited applicability for complex systems. This technique lacks the dynamic adaptability to target potential energy landscapes with multiple local minima separated by large energy barriers often encountered in processes such as biomolecular self-assembly.

Metadynamics (MetaD), an enhanced sampling technique developed by Laio and Parrinello in 2002, (Laio and Parrinello 2002) overcomes these limitations by introducing a history-dependent biasing potential, which flattens the free energy surface and facilitates the crossing of energy barriers. The authors have reasoned that these traditional methods often struggle to escape local minima in the free energy surface due to high energy barriers, limiting their effectiveness in studying processes like protein folding. Metadynamics enables cost-efficient exploration of rare events and thermodynamic properties in systems inaccessible through standard MD simulations.

Metadynamics addresses this by employing coarse-grained, non-Markovian dynamics in a space defined by collective coordinates. A key feature of this method is the inclusion of a bias potential, typically implemented as a sum of Gaussian functions. These Gaussians "fill" the free-energy wells over time, flattening the landscape and enabling the system to escape local minima. This iterative process not only facilitates the exploration of the free energy surface but also allows its quantitative reconstruction as a function of the chosen collective coordinate, providing insights into the thermodynamics and kinetics of the system under investigation.

The method's fundamental principles, assumptions, and procedures are outlined for the 2002 study (Laio and Parrinello 2002), later recognized as the standard metadynamics approach (Standard MetaD). Bussi and Laio (2020) summarized the foundational steps, noting that the success of metadynamics depends significantly on the selection of the CVs and biasing. This methodology assumes a finite set of n relevant CVs, $s_i$, to describe the system's free energy surface, F(s). Forces guiding the exploration of the free energy surface are defined as:

$$F_i = -\partial F / \partial s_i \quad (13)$$

To estimate these forces, an ensemble of P replicas is generated, each constrained to a specific $s_i$ value using Lagrange multipliers. The average force is evaluated as $F_i = \langle \lambda_i \rangle$, where $\lambda_i$ are the multipliers.

The exploration employs a coarse-grained dynamics approach, using the evolution equation:

$$\sigma_i^{t+1} = \sigma_i^t + \delta\sigma(\phi_i^t / |\phi^t|) \quad (14)$$

where $\sigma_i = s_i/\Delta s_i$ are scaled variables, $\phi_i = F_i \Delta s_i$ are scaled forces, and $\delta\sigma$ is a stepping parameter. This equation represents the steepest descent in the free energy surface. To enhance sampling, the history-dependent bias is introduced by replacing the forces $F_i$ with $F_i + \sum_k W^* \exp(-|s-s_k|^2/(2\delta s^2))$, based on:

$$\phi_i \to \phi_i - \frac{\partial}{\partial \sigma_i} W \sum_{t' \leq t} \prod_i e^{-\frac{|\sigma_i - \sigma_i^{t'}|^2}{2\delta\sigma^2}} \quad (15)$$

where W and δs are respectively the height and width of Gaussian functions. The Gaussians discourage revisiting previously sampled regions and enabling transitions between them. For large t (and especially if the width of the Gaussians is sufficiently small compared to the length of a typical variation of V), the free energy surface can be reconstructed as $F(s) \approx -\sum_k W^* \exp(-|s-s_k|^2/(2\delta s^2))$, where:

$$-\sum_{t' \leq t} W e^{-\frac{|\sigma - \sigma t'|^2}{2\delta\sigma^2}} \to V(s) \quad (16)$$

The efficiency of the method scales with $(1/\delta\sigma)^n$, and judicious choices of $\Delta s_i$, $W$, and $\delta\sigma$ balance accuracy and efficiency. The Gaussian height W satisfies: $(W/\delta\sigma)e^{-1/2}=\alpha\langle F\rangle^{1/2}$, with $\alpha<1$, typically around 0.5 from their test systems.

It is also noted that while their early examples focus on biomolecules, they have yet to highlight the application of metadynamics in assembly processes. However, the versatility of metadynamics is reflected in its variability of scaling different bias potentials to different scientific problems. Therefore, they suggested adaptive parameter selection for future improvements to better fit local free energy surface features.

Later research expanded metadynamics subcategories to refine its performance and expand its applicability. Gao et al. (2019) categorized metadynamics as CV-based sampling that employs predefined CVs to guide the simulation. The equilibrium distribution, $p_0(s)$, and the free energy, $F(s)$, are derived from the Boltzmann distribution, allowing the system to sample less probable states effectively. Therefore, some of the most prominent derivatives of metadynamics have been focusing on enhancing sampling efficiency.

Bias-exchange metadynamics (BE-MetaD) extends the method to high-dimensional systems by employing parallel simulations with different CVs, broadening the sampled configuration space. (Piana and Laio, 2007) The method utilizes multiple replicas running at the same temperature, each biased by a different CV. These replicas periodically exchange conformations and enhance diffusion in the CV space within the system trajectory. The results are not direct multidimensional free energy surfaces but rather low-dimensional projections along different CVs.

Well-tempered metadynamics (WT-MetaD) was also developed (Barducci et al., 2008) to improve stability by reducing artifacts and oversampling. This method gradually reduces the height of the Gaussian functions as the bias potential grows, ensuring convergence of $V(s)$. The height of the Gaussian functions decreases as $\omega(t)=\omega/(1+\beta\gamma^{-1}V(s,t))$, where $\gamma$ is the bias factor. This refinement allows WT-MetaD to reconstruct accurate free energy surfaces even for systems with high energy barriers.

Multiple walkers metadynamics (MW-MetaD) employs parallel trajectories, or "walkers," that share a common CV and deposit Gaussian functions into a shared bias potential. This parallelization accelerates exploration by distributing sampling efforts similar to REMD (Williams-Noonan et al. 2023).

Variationally enhanced sampling (VES) introduces a functional approach to optimize the bias potential (Valsson and Parrinello 2014). The functional $\Omega[V(s)]$, often defined as the Kullback-Leibler divergence, minimizes the difference between the target distribution and the actual distribution. The bias potential in VES is represented as $V(s)=\sum_k a_k\phi_k(s)$, where $\phi_k(s)$ are basis functions and $a_k$ are parameters optimized using stochastic gradient descent. This method allows dynamic optimization of the bias potential, bridging enhanced sampling methods with machine learning techniques.

Other developments, such as parallel-tempered metadynamics and on-the-fly parameter optimization, integrate multicanonical simulations to achieve a uniform sampling of states across energy levels.

Parallel-tempered metadynamics (PT-MetaD) combines replica exchange Monte Carlo with adaptive biasing in MetaD to enhance sampling efficiency in Molecular Dynamics simulations (Bussi et al. 2006). Gaussians are deposited along selected CVs at a frequency $1/\tau_G$, progressively flattening the free-energy landscape:

$$V_i(s; t) = V_i(s; t - \delta t) + w_i e^{-\frac{(s - s_i(t))^2}{2\sigma_G^2}} \quad (17)$$

The Gaussian height $w_i$ is scaled with temperature, typically following $w_i \propto T_i$. This ensures that higher-temperature replicas rapidly explore the free-energy surface, allowing low-temperature replicas to decorrelate faster. The CVs used for biasing include the gyration radius of the backbone and the number of hydrogen bonds. A geometric temperature distribution is also employed for constant acceptance ratios across replicas to maximize exchange efficiency.

Expanding on BE-MetaD, the development of Parallel Bias Metadynamics (PB-MetaD) also utilizes parallelization to apply multiple bias potentials, following the WT-MetaD form, across different CVs. PB-MetaD introduces a discrete variable η that toggles bias potentials corresponding to the active CV, to prevent bias contamination. By weighting the deposited Gaussians by the conditional probability P(η | R), it allows straightforward reweighting to recover high-dimensional free-energy landscapes (Pfaendtner and Bonomi 2015).

Comparing these methods reveals distinct advantages tailored to different simulation challenges. To address bias convergence issues for Standard MetaD, WT-MetaD focuses on ensuring reliable free energy reconstruction, while MW-MetaD and BE-MetaD enhance efficiency through parallelization. Later developments of MetaD — VES, On-the-fly parameter optimization, PT-MetaD and PB-MetaD — reduce non-ergodic sampling through adaptive biasing, where bias potentials evolve based on system response, showing joint interest in the field to balance both aspects and introducing prospective machine learning-based optimization, offering more powerful tools for modern simulations. (Invernizzi and Parrinello 2022; Zerze et al. 2023)

**Surrogate Models in Enhanced Sampling of Self-assembly Process using ML Techniques**

Studying biomolecular self-assembly *in silico* poses numerous challenges such as estimating complex free energy landscapes, slow transitions between different metastable states, and sampling the long time scale of a self-assembly process. The two major limitations in molecular simulations as highlighted by Karplus et al (Karplus and Petsko 1990), are the approximations in the potential energy functions and the lengths of the simulations. The first introduces systematic errors and the second statistical errors, respectively. The first can be addressed by developing accurate force fields informed by quantum mechanical calculations or experiments for various molecular chemical species. The second requires the use of novel enhanced sampling techniques in order to capture and simplify the complex rugged free energy landscape of proteins and other macromolecules. ML models provide an organic framework for high-fidelity enhanced sampling methods as most of the problems in ML also overlap with approaches in MD under different names like dimensionality reduction, depositing biasing potential, and capturing tractable, intractable probability densities (Mehdi, Smith et al. 2024).

Simulations of long timescale phenomena like nucleation, and self-assembly accumulate overhead sampling costs of crossing metastable transition barriers, and hence require strategies to accurately describe reaction RCs or CVs for biasing potentials. Various deep learning models have been used to leverage artificial neural networks (ANN) in order to learn the optimal submanifold of the free energy landscape by discovery of CVs such that only dominant metastable states or slow dynamics are captured.

Based on Markov State Model (MSM), Mardt *et al.* (Mardt, Pasquali et al. 2018) developed a deep learning framework utilizing a variational approach for the Markov process (*VAMPnet*) which combines all the tasks of MSM, featurization of short MD trajectories, dimensionality reduction of features and clustering into microstates. VAMPnet relies on time-lagged configurational MD data as input in two deep networks (i.e., Siamese neural network)(Figure 4) combined by maximizing VAMP-2 score (eq 19) following the optimization of latent variable sets related to each other by the Koopman operator (eq 18).

$$E[\chi_1(x_{t+\tau})] \approx K^T E[\chi_0(x_t)] \tag{18}$$

$$\widehat{R}_2[\chi_1, \chi_0] = \left\| C_{00}^{-1/2} C_{01} C_{11}^{-1/2} \right\| \tag{19}$$

where $C_{00} = E_t[\chi_0(x_t)\chi_0(x_t)]$, $C_{01} = E_t[\chi_0(x_t)\chi_1(x_{t+\tau})]$, $C_{11} = E_{t+\tau}[\chi_1(x_{t+\tau})\chi_1(x_{t+\tau})]$

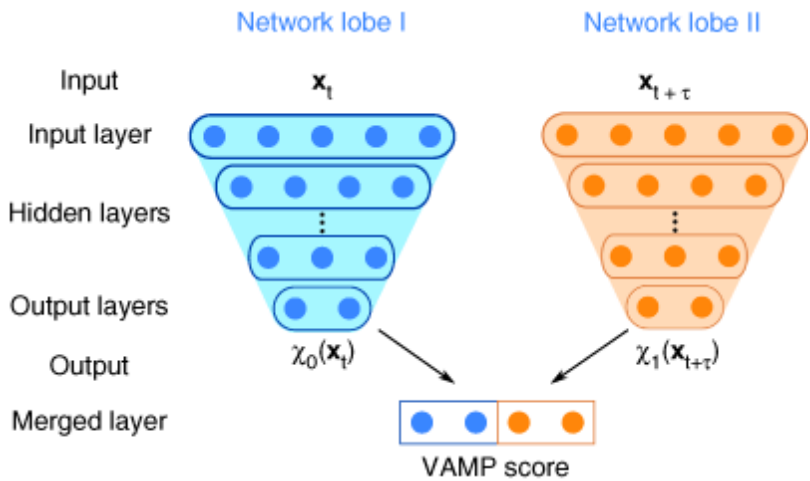

Figure 4: Siamese neural network of VAMnet containing two sub-networks with same weights (Mardt, Pasquali et al. 2018).

VAMPnets can be used to identify CVs corresponding to the slow modes of self-assembly by appropriately selecting a lag time, which can be determined based on the autocorrelation time decay of relevant order parameters, such as cluster size, radial distribution functions, or pairwise distances. However, VAMPnets are utilized to learn discrete metastable stable states as output and since meaningful enhanced sampling requires continuous, interpretable RCs, this limits its application. Chen et al. (Chen, Sidky et al. 2019) proposed state-free VAMPnet (SRV) where the goal is not to partition the state space but to learn a continuous non-linear function of the input data that serves as a basis for approximating eigenvalues of transfer operator within a variational approach for conformational (VAC) dynamic framework. SRV consists of a Siamese neural network where a pair of time-lagged configurations are fed into the same architecture having the same weights. The non-linear mapping of latent variables from the lobes is plugged in the generalized eigenvalue equation of the transfer operator and overall the model is trained to minimize the loss function $L = \Sigma_i\, g(\widetilde{\lambda_i})$, where $\lambda_i$ is eigenvalue and $g$ is a monotonically decreasing function of the eigenvalue.

Another approach utilizing the VAC dynamics is Deep-TICA (Deep-Time Lagged Independent Component Analysis) by Bonati et al (Bonati, Piccini et al. 2021), which enables efficient sampling of rare events, such as molecular conformational changes and crystallization, enhancing simulation convergence for studying complex phenomena. In TICA, the eigenfunction of the transfer operator is constructed as a linear combination of descriptors of the system (eq 20). A detailed balance condition makes the transfer operator a self-adjoint matrix. The matrix elements are computed by time auto-correlation of the descriptors in a generalized eigenvalue problem (eq 21).

$$\widetilde{\psi_i}(R) = \Sigma_{j=1}\, \alpha_{ij}\, d_j\,(R) \qquad (20)$$
$$C(\tau)\, \alpha_i = \widetilde{\lambda}\, C(0)\, \alpha_i \qquad (21)$$

where $C_{ij}(\tau) = <\, d_i\,(R_t\,)\, d_j(R_{t+\tau}\,) >$
$C_{ij}(0) = <\, d_i\,(R_t\,)\, d_j\,(R_t) >$

Deep-TICA consists of ANN that takes time-lagged descriptors as input and non-linearizes the linear TICA method by outputting the autocorrelation coefficient as a function of latent variables which is further plugged in the generalized eigenvalue equation. The whole problem becomes maximization of eigenvalues such that errors are back propagated using a loss function, $L = -\Sigma_i\, \widetilde{\lambda}\,(\theta)$, where $\theta$ represents a set of parameters, namely weights and biases of ANN. The Deep-TICA method is instrumental in learning slow modes from a biased timescale starting from suboptimal CVs, nevertheless combining both the trajectories, one that of the initial descriptive RCs and the other corresponding to the learned RCs, posing a non-trivial challenge.

Another unsupervised nonlinear reduction technique for uncovering CVs is diffusion maps (dMaps), which compute the embedding of high-dimensional data into low-dimensional space whose coordinates can be computed from eigenvectors and eigenvalues of a diffusion operator on the data (Coifman, Lafon et al. 2005) (Ferguson, Panagiotopoulos et al. 2011). Ferguson et al (Long and Ferguson 2014) employed dMap to study the kinetics and thermodynamics of self-assembly pathways of patchy colloidal particles using Brownian dynamics. In order to implement dMap on simulation data, a suitable cluster distance matrix has to be defined to measure the kinetic proximity of clusters to be formed in terms of cluster similarity. Since such a metric must be rotational and permutation invariant and give insights into the local structure of the clusters, the cluster similarity problem was transformed into a graph similarity problem using a spectral graph matching algorithm. Unlike other ANN models, dMap uses a Gaussian kernel to generate a right-stochastic Markov matrix from the symmetric matrix of pairwise cluster similarities (eq 22)

$$A_{ij} = exp\left(-\frac{d^2_{ij}}{2\varepsilon}\right) \qquad (22)$$

Where $d_{ij}$ is cluster distance or cluster similarity metric between ith and jth cluster and $\varepsilon$ is the bandwidth of gaussian. Following row normalization of A matrix, it can transformed into a Markov matrix M as shown below (eq 23-24)

$$D_{ij} = \Sigma^N_{k=1} A_{ik}, \quad if\ i = j \qquad (23)$$
$$, otherwise$$

$$M = D^{-1} A \qquad (24)$$

The eigenvectors of the M matrix are discrete estimations of the eigenfunctions of the Fokker-Plank operator, describing a diffusion process over the data. The collective modes above the spectral gap of eigenvalues steer the long-time evolution of the system. Projecting the i[th] cluster along the i[th] component of these eigenvectors gives the slow, dominant mode dynamics driving self-assembly. However, dMaps suffer from two main limitations (Sidky, Chen et al. 2020), first is the assumption that diffusive dynamics over the high-dimensional data may or may not be a good approximation of the true molecular dynamics. The second is that the dMaps don't define an explicit mapping from atomic coordinates to the low-dimensional CVs, rendering it incompatible with enhanced sampling methods like umbrella sampling or metadynamics which require the gradients of atomic coordinates with respect to atomic coordinates.

Some of the dMap-based enhanced sampling methods are diffusion-map-directed MD (DM-d-DM) (Rohrdanz, Zheng et al. 2011) and intrinsic map dynamics (iMapD) (Chiavazzo, Covino et al. 2017). DM-d-DM method enhances sampling at the frontier of explored space, initiating an unbiased simulation rather than imposing artificial bias. The local scaling of the gaussian kernel width $\varepsilon$ around a certain configuration is performed through multidimensional scaling that reduces high dimensional configurational space to a "locally flat" low dimensional space (Rohrdanz, Zheng et al. 2011). The parameter $\varepsilon$ determines how noise in a highly variable conformational density is reduced. If $\varepsilon$ is too small the results will be all noise in a "locally flat" region and if $\varepsilon$ is too large, regions will be artificially flat. In iMapD, similar to DM-d-DM, the short simulation runs are embedded using dMaps, but the boundary of the explored space is extended outwards by a certain amount using local principal component analysis (PCA). The new unbiased simulation is produced from the extended region; in this iMap method, it is said to tunnel through the free energy landscape. Though both of these methods can approximate true MD over the energy barrier by not letting molecules tumble downward from the energy barrier, the resulting CVs are still discontinuous and non-differential. This can be overcome using ANN-based diffusion net (DNETS) (Mishne, Shaham et al. 2019), which trains an ANN encoder to learn functional maps from atomic coordinates to a low-dimensional dMap.

Self-assembly conventionally studied in only thermodynamic terms is insufficient to give a full statistical description of the major underlying kinetic pathways (Conti and Cecchini 2016) (Conti, Del Rosso et al. 2016) (Palma, Cecchini et al. 2012). Kinetic network models (KNMs) (Liu, Qiu et al. 2022) can be deployed to study the kinetic control of self-assembly by combining many MD simulation trajectories like the Markov state model. However, KNMs suffer from two major challenges for their applications in the process of self-assembly (Liu, Xue et al. 2023). First, the selected features that are descriptive of self-assembly should be invariant to permutations and rotational symmetries in the system. Second, self-assembly is typically an energetically downhill process in a free energy landscape that can involve the trajectory of monomer aggregation from out-of-equilibrium states with insufficient sampling of the dissociation of monomers. This is opposed to the way KNMs extract slow CVs by imposing detailed, balanced conditions. GraphVAMPnet (Liu, Xue et al. 2023) resolves these issues by combining the workflow of high-dimensional dynamic graph embedding of the molecules and VAMPnet operating on the time-lagged graph-embedded conformations of molecules as inputs. The graph representation of molecules in GraphVAMPnet is based on the SchNet framework (Schütt, Sauceda et al. 2018) with the notion that the particles with the same permutation and rotational symmetries should belong to the same node embedding. The edges between the nodes are defined by an extensible pairwise gaussian vector (Eq(8)) characterized by a cut-off distance between pairwise particles in the nearest neighbor region.

$$e_{ij}^{\ k} = exp\left(-\frac{(d_{ij} - c_k)}{\sigma^2}\right) \tag{25}$$

where $e_{ij}^{\ k}$ is the *k*th element of edge vector $e_{ij}$, $d_{ij}$ is the pairwise distance between node i and node j, and $c_k$ is the center of the expansion.

All node vectors and edge vectors in the graph at a particular layer are mapped into the same dimension by dense layers. These are further input into the attention-weighted continuous convolution layer in which the local interaction of each particle with its surroundings is obtained through attention weights. Attention weight is representative of the relative importance of all adjacent particles of a central particle in a convolution. The local interactions on particles obtained from this layer are projected back and added to the original node vectors while the edge vectors remain the same. The time-lagged conformations sent through this graph neural network are passed onto the Siamese neural network (i.e., weights in two ANNs are the same) of the VAMPnet. The set of lagged features is transformed in the hidden layers to optimize the singular value decomposition problem using the Koopman operator. The optimally approximated leading singular function of the Koopman operator corresponds to the slowest CV of the reduced kinetic space. While the graph neural network ensures that the system's local and global interaction environment is captured following permutation and rotational invariance, VAMpnet allows for partial reversibility of the formation of self-assemblies, exhibiting the true dynamics of the system.

The cooperative, collective interaction steering self-assembly of molecules can also be captured by Transition Path Sampling (TPS). The transition paths (TPs) are special trajectory segments that capture the rare event reorganization going through the infrequent, rapid transitions between different metastable states. Since TPs are stochastic in nature, hence they require a probabilistic framework using the committor function. A committor function $p_s(x)$ represents the probability that a trajectory enters state S first, with S = A or B, where x is a vector feature of the starting point X in the configuration space. The committor to a target state reports on the progress of the reaction and predicts the trajectory in a Markovian way, thus making it an ideal RC. Jung et al. (Jung, Covino et al. 2023) designed an iterative algorithm to learn the committor of rare events using a neural network in a way similar to reinforcement learning. The committor probability is encoded in the dense layers of the neural network, where the transition path is progressively weighed in favor of the reactant from many attempts to maximize the committors along trajectories by using negative log-likelihood of it as a loss function. The neural network architecture is pyramidal so as to arrange the features in a resolution hierarchy, from the highest possible resolution of the cartesian coordinates of atomic positions to a singular quantity of committor. The mechanism has been successfully applied to study ion association in solution and protein-membrane assembly (Jung, Covino et al. 2023).

The use of autoencoder architecture offers a promising role in the construction of low-dimensional free energy where a somewhat complete separation of timescales can be achieved. These architectures consist of encoder-decoder multilayer perceptrons in which the encoder section compresses the input data to a lower dimensional space also known as latent space and the decoder section reconstructs the reduced data to its original size. The objective of the learning process is to minimize the deviation between the input and reconstructed data. The input data as structural descriptors of the molecular system must abide by simple symmetries like invariance with respect to translation, rotation, and permutation. Appeldorn et al (Appeldorn, Lemcke et al. 2022) employed a neural network autoencoder called EncoderMap (Lemke and Peter 2019) in order to study the self-assembly of two complementary single-stranded DNA. They used complementary base pair distance as the input structural features, which are mapped onto a latent space in the neural network, and an MSM is constructed in this space. This allows for the inspection of the latent space coordinates as order parameters.

## Section 4 - Applications

**Application of Enhanced Sampling Techniques to Different Biomolecular Systems**

### *(1) Peptides*
REMD has been used in numerous studies concerned with the self-assembly of short peptide chains encompassing between 4 and 30 amino acid residues. Most of these studies are related to amyloids or amyloidogenic fragments (Anand 2008; Baumketner 2005; Cecchini 2004; Gee 2011; Jeon 2012; Martel 2017; Takeda 2008), chosen for their relevance to human diseases including Alzheimer's Disease (AD). REMD has also been used to examine phenylalanine-containing peptides which form self-assembled peptide nanostructures (Jeon 2014; Rissanou 2013; Tamamis 2009; Xiong 2019); transmembrane and membrane-bound peptides (Kokubo 2004; Martel 2017; Urano 2015); the formation of peptide dimers and oligomers (Baumketner 2005; Gee 2011; Jeon 2012; Soto 2010; Takedo 2008; Wolf 2008), and protein-ligand binding (Lockhart 2023).

As previously noted, the challenge of identifying the forces driving the self-assembly of biomolecules, such as peptides, requires estimating free energy differences using either a single umbrella sampling method or combining methods such as umbrella sampling with REMD. Yu and Schatz applied umbrella sampling in conjunction with targeted MD to understand the factors which drive the self-assembly of peptide amphiphiles to form micelles (Yu and Schatz 2013). The study determined the driving force to be enthalpic, in particular electrostatic interactions followed by van der Waals interactions. The mechanism of self-assembly was determined from the potential of mean force (PMF) (Yu and Scatz 2013).

Wolf et. al. 2008 investigated the mechanism underlying the self-assembly of peptides into amyloid fibrils, and found that the final amyloid fibril composition was governed by thermodynamics. The study reported a first order phase transition, which was linked to the formation of amyloid fibrils. The authors used the free energy landscape to predict the final composition of the self-assembled amyloid fibrils and compare their findings to experiments. In this study, REMD was used alongside umbrella sampling with a harmonic bias to calculate the PMF, and was observed to provide a tenfold increase to speed compared to traditional umbrella sampling.

Jeon and Shell (2014) examined the self-assembly of cyclo-diphenylalanine (cyclo-FF) in vacuum and compared it to the assembly of linear FF peptides in water. The study applied umbrella sampling replica-exchange MD (U-REMD) with harmonic constraints to study the interaction between cyclo-FF molecules and determine the PMF. Cyclo-FF molecules assembled to form a ladder-like structure due to electrostatic and van der Waals interactions in the backbone. The authors noted that this was different from the assembly of linear FF in water where the driving forces were arising from the hydrophobic effect and electrostatic interactions of the sidechains.

Laio et al. (2012) used BE-MetaD to explore the aggregation process of 18 chains of 8-valine (VAL8) in a disordered state to an ordered β-sheet structure. The BE-MetaD simulation was run for 4340 ns, with 8 replicas biased by 8 different CVs, each corresponding to different features of amyloid aggregation (such as parallel/antiparallel β-strands and packing of β-sheets). Convergence of the bias potential was monitored, and free energy estimates were derived using WHAM. The system explored hundreds of configurations. The results revealed distinct free energy minima for structures rich in antiparallel β-sheets and those with more parallel β-strands. By highlighting a complex, non-trivial aggregation pathway, this method provided valuable insights into the formation of amyloid-like structures and the influence of peptide sequences on the dynamics of aggregation, such as the transition from antiparallel to parallel β-sheets.

More recently, Pfaendtner and Sampath (2019) employed Parallel-Tempering Metadynamics in the Well-Tempered Ensemble (PTMetaD-WTE) to investigate the binding behavior of the LKα14 peptide on crystalline and amorphous silica surfaces. The assumptions underlying the use of PTMetaD-WTE in this research include the hypothesis that peptide binding to the surface is governed by electrostatic interactions and that the peptide adopts a helical structure upon adsorption. By focusing on CVs related to the peptide's interaction with the surface and its conformation, the method allowed for the efficient exploration of the peptide's conformational landscape and its binding affinity to the silica surface. The use of PTMetaD-WTE facilitated the analysis of peptide binding on both crystalline and amorphous surfaces, highlighting the differences in binding strength and conformational diversity between the two surface types.

## *(2) Proteins*

Formation of quaternary structure in protein complexes has been studied with REMD (Siradzan 2012). The long time scales associated with spontaneous self-assembly require the simulated systems to be placed under constraints – such as fixing the COM distance between two molecules – to allow examination of the conformation ensembles and energetics associated with close contact between particles. The analysis of constrained or biased MD simulations can provide key information about the processes underlying self-assembly. Constrained systems are often used to characterize a PMF, binding free energy, or another energy surface. Such energy surfaces may be used for developing coarse-grained potentials.

Yuan et al. (2020) employed metadynamics simulations to investigate the dissociation mechanisms of AG10 and tafamidis from the thyroxine-binding sites of the TTR tetramer. By analyzing the dissociation pathways, the study identified key interactions, such as salt bridges and hydrogen bonds, crucial for the dissociation of AG10. Whereas the dissociation of tafamidis involved more intricate interactions, including cation-π and hydrogen bond formations. The simulations revealed two major energy wells for the drug dissociation pathways, showing that tafamidis has a stronger binding affinity. AG10 exhibits more contacts with the target protein, while tafamidis exhibits a more complex multistep unbinding process involving three energy wells.

Well-tempered metadynamics (WT-MetaD) was used by Xu et al. (2022) to study the binding mechanism between the supramolecular tweezer CLR01 and the lysine residues on the 14-3-3σ protein. This was done by enhancing the sampling of binding conformations and calculating binding free energies. The study employed WT-MetaD to explore the thermodynamics and kinetics of this interaction, with two CVs chosen to track the distance and coordination number between the tweezer and lysine residues. This method allowed for the construction of a comprehensive free energy surface that helped identify binding sites with different affinities for the tweezer. Among the 17 lysine residues on 14-3-3σ, 8 sites were found to form inclusion complexes, with some exhibiting strong binding (K214) and others showing moderate binding. The simulation results, which aligned well with experimental measurements, revealed that spatial hindrance around the lysine residues was the key factor influencing binding.

Salvalaglio et al. (2022) used Parallel-Tempering Metadynamics with Well-Tempered Ensemble (PTMetaD-WTE), to enhance the exploration of DHH1N's conformational ensemble by accelerating sampling of its free energy landscape. Through biased simulations using CVs such as Cα–Cα and Cγ–Cγ contacts, backbone hydrogen bonds, and structural parameters, PTMetaD-WTE effectively mapped the protein's configurational states, revealing the intrinsic disorder of DHH1N. The analysis of the one-dimensional free energy profiles highlighted a primary global minimum corresponding to a largely disordered state, consistent across CHARMM36m and CHARMM22* force fields. Two-dimensional free energy surfaces further confirmed a weakly funneled landscape with distinct metastable states, providing insight into the protein's conformational flexibility. Complementary BE-MetaD simulations reinforced these findings by expanding the sampling of residue-residue interactions and secondary structure motifs, particularly revealing β-strand formation linked to specific residue contacts in CHARMM36m. Overall, PTMetaD-WTE helped detail a statistically robust representation of DHH1N's free energy landscape, capturing its disordered nature while elucidating subtle force-field-dependent structural propensities.

### *(3) Polymers*
Umbrella sampling with harmonic potentials was utilized to calculate the free energy between different aggregates of an amphiphile and polyethylene glycol (PEG) polymer (Brunel et. al. 2019). The authors found that the inclusion of PEG side chains created strong repulsive barriers, resulting in the aggregates to transition from spheres to larger vesicles. Weaker repulsive barriers led to formation of smaller vesicles and tubular nanostructures.

In Mu et al. (2017), well-tempered metadynamics (WT-MetaD) significantly enhanced the exploration of the free energy surface of TXB-LII binding, overcoming the limitations of unbiased MD simulations. Due to the high flexibility of both TXB and LII, the binding landscape is rugged, characterized by numerous local minima arising from the presence of highly polar residues along with the polar ring motif in TXB. Unbiased MD simulations, despite being run for 200 ns across three independent repeats, failed to identify a stable binding mode, with configurations often getting trapped in local minima for extended periods. The study employs PTMetaD-WTE to efficiently explore the free energy landscape of GB1 dimerization under different crowding conditions by biasing key RCs. The convergence of the PTMetaD-WTE simulation was verified through the diffusion of the CV and decay of the Gaussian height at 300 K, indicating that the enhanced sampling successfully overcame energy barriers and provided a more complete free energy surface. The simulations reveal that lysozyme crowders destabilize GB1 dimers, particularly the side-by-side configuration, by shifting the free energy minimum to a state where crowders disrupt native interactions. Comparative simulations with truncated LII analogs (C15P and C15PP) demonstrated that the pyrophosphate and GlcNAc-1 groups were essential for stable TXB binding, reinforcing the importance of electrostatic complementarity. Overall, PTMetaD-WTE not only facilitated the identification of stable TXB-LII binding modes but also provided mechanistic insights into the molecular determinants of affinity and specificity.

Hirokawa et al. (2018) used WT-MetaD simulations to understand the binding kinetics of amantadine and adamantyl bromothiophene to M2 channels (S31 and N31 mutants). Metadynamics identified the binding of amantadine to the S31 M2 channel as characterized by three steps (L1, TS, L2), together with the free energy profiles to map these transitions. Similar insights were gained for adamantyl bromothiophene, which showed more complex binding dynamics. For adamantyl bromothiophene, the metadynamics simulations also helped elucidate its dual binding mode in S31 and N31 M2 channels, which could not be captured by conventional methods. The broad, funnel-shaped energy profiles indicated differences in binding kinetics between the two M2 mutants.

In Parrinello et al. (2013), metadynamics played a critical role in understanding the dimerization and trimerization processes of fibritin by probing protein-protein interactions. For the dimerization process, metadynamics was used to apply bias to two configurational CVs: the distance between the COM of the monomers and the number of specific monomer–monomer contacts. The analysis of the dimer's structural ensemble suggested that various configurations, including an almost antiparallel dimer arrangement, were thermodynamically favored, even in the absence of a third monomer. For the trimerization process, the PTMetaD-WTE approach was also applied, using similar biasing CVs. The results revealed a funnel-shaped free-energy landscape, strongly biased toward the formation of the experimental trimeric structure, with minimal structural heterogeneity compared to the dimerization process. The trimer's binding free energy closely matched experimental values.

From the studies of Eslami and Müller-Plathe (2024), adaptive-numerical-bias metadynamics was employed to investigate the self-assembly of triblock Janus nanoparticles into open and close-packed 3D colloidal crystals. This method progressively adds adaptive biasing potentials to the system's unbiased potential energy, enabling efficient traversal of high-energy barriers between phases. A key assumption in this approach is that suitable order parameters can distinguish different phases, guiding the simulation toward relevant regions of the free-energy surface. The use of bond-orientational order parameters based on spherical harmonics, allowed for precise characterization of local crystalline order, distinguishable between phases. The metadynamics simulations revealed that the free-energy barriers for nucleation of open lattices are generally lower than those for denser lattices, a finding attributed to the lower density mismatch between the liquid and open crystalline phases. By iteratively refining the biasing potential, the metadynamics approach successfully mapped the free-energy basins of different phases and identified transition pathways, offering valuable insights into the nucleation and growth of colloidal crystals.

### *(4) Nucleic Acids*
In the research of the 22-nt monomolecular DNA G4 sequence folding process, Moraca et al. (2020) employed Well-Tempered Metadynamics (WT-MetaD) simulations to overcome the time-scale limitations inherent in molecular simulations and to accelerate the exploration of rare events during the formation of the G4 structure. WT-MetaD allowed the identification of significant metastable states, which are crucial for understanding the folding process of G4 DNA. The simulations focused on two key CVs: the coordination of Hoogsteen hydrogen bonds within the G-tetrads and the π-π stacking interactions among the guanines of the G-triplets. WT-MetaD was instrumental in revealing an ensemble of conformations that had previously been undetectable, particularly those involving partial G-tetrad planes and varying degrees of strand slippage. The simulation results showed that vertical slippage of G-triplets was a key mechanism in G4 folding, with two adjacent G-triplets exhibiting a higher probability of slippage than opposites. The study's findings also supported the hypothesis that the formation of G-triad intermediates, involving slippage and incomplete G-tetrads, could play a crucial role in the early and late stages of G4 formation and disruption. By reweighting the sampling data from the WT-MetaD simulation, the authors were able to construct a free energy surface that highlighted the stability and probability of various G4 conformations.

Ferrarini et al. (2023) employed WT-MetaD to obtain a detailed description of the free energy landscape of $Zn^{2+}$ binding to ATP for the self-assembly of materials exhibiting non-equilibrium behaviors, which involved identifying distinct coordination modes between $Zn^{2+}$ and phosphate oxygens in the ATP triphosphate chain. By using WT-MetaD, the simulations provided insight into the most stable $Zn^{2+}$–ATP binding mode and were essential in understanding how the conformational flexibility of the triphosphate chain influences the stability of the binding. This method enabled a robust analysis of free energy differences and coordination geometries, thereby refining the understanding of $Zn^{2+}$ coordination in ATP and informing future coarse-grained simulations examining assembly/disassembly processes.

In Wang et al. (2024), WT-MetaD played a central role in unraveling the pH-dependent assembly mechanism of the TLR3×4/dsRNA signaling complex by enabling in-depth exploration of its free energy landscape. To incorporate environmental pH effects, the study combined metadynamics with constant pH MD (CpHMD), enabling the investigation of dynamics underlying pH-dependent assembly. This hybrid approach was crucial for analyzing the protonation states of titratable residues. Four CpHMD-metadynamics simulations were conducted based on pH values, focusing on combinations of RCs. Convergence was confirmed using block-analysis techniques and the plateauing of the PMF curves, highlighting the robustness of the methodology. The metadynamics simulations provided critical insights into the stability of the TLR3×4/dsRNA signaling complex. The results demonstrated that the complex remains stable under acidic conditions (pH 6.0) but disassembles in basic environments (pH 7.4).

In Zerze et al. (2023), two advanced sampling techniques, MM-OPES and PTWTE-WTM, were compared for atomistic RNA simulations, specifically focusing on the GAGA tetraloop. Both methods aimed to explore the free energy landscape of the tetraloop by investigating its various folding states: unfolded, folded, and misfolded. Specifically, the MM-OPES technique utilized multiple thermodynamic temperatures to balance sampling efficiency with accuracy. One of the key assumptions of MM-OPES is that high temperatures aid in improving ergodicity, allowing for more efficient exploration of the system's conformational space, a concept borrowed from parallel tempering. The study found that temperature sets with a minimum temperature near 300 K and a maximum temperature above 350 K were most effective for this system, with lower maximum temperatures failing to explore critical regions of the free energy landscape. MM-OPES results showed high accuracy in reproducing the free energy landscape obtained from PTWTE-WTM simulations for many temperature sets. Whereas PTWTE-WTM demonstrated better sampling at the folded state. The authors highlighted that the reduced cost of MM-OPES, approximately four times less computationally expensive than PTWTE-WTM while yielding similar results, further demonstrated its potential for practical applications in biomolecular simulations.

### (5) Supramolecules

Standard metadynamics has been employed for predicting free energy surfaces in situations like ligand docking. Gervasio et al. (2005) noted that the adaptation of metadynamics for docking included selecting metavariables that could balance efficiency with accuracy, such as ligand-receptor distance and angles related to their orientation. The metadynamics trajectory was used to test various docking configurations, refine ligand positioning, and compute the binding free energy ($\Delta G\_binding$). It is found that the method reduced the need for high-dimensional search spaces, which are typically computationally expensive. These results were normally compared with experimental data with satisfying alignment.

In Pavan et al. (2021), WT-MetaD simulations were used to investigate the dynamics of nanoparticle (NP) unbinding and escape from surface receptors. WT-MetaD improved the understanding of the free energy barriers and characteristic time scales associated with NP detachment, in both fine-grained coarse-grained (fCG) models and all-atom simulations. By selecting the number of contacts between NP beads and receptor beads as the CV, the researchers applied a bias to overcome the free energy barriers associated with NP detachment. The simulations were run with a bias factor of 10 and Gaussian kernels. The characteristic time scale for NP unbinding was obtained by fitting the unbiased transition times to a Poissonian distribution. This provided insights into the dynamics of NP-receptor interactions, with a specific focus on the rate of NP detachment and the influence of receptor density on the unbinding process. Further refinement of the approach was achieved by running MW-MetaD simulations, which allowed for the parallel exploration of the NP behavior on surfaces with varying receptor densities. The results from these simulations were consistent regardless of whether long-range electrostatics were treated using Particle Mesh Ewald summation, demonstrating the robustness of the methodology in modeling NP chemotactic behavior.

Biswas et. al (2023) used the Parallel Tempering Metadynamics with Well-Tempered Ensemble (PTMetaD-WTE) approach to investigate the dimerization of GB1 monomers under varying conditions of molecular crowding. PTMetaD-WTE introduced efficient exploration of the conformational space associated with protein-protein interactions. The study assumed that the relevant RCs are sufficient descriptors of the dimerization process, allowing the free energy landscape to be reconstructed, and providing insights into the thermodynamic favorability of dimer formation. The results demonstrated that, in the absence of crowders, both the side-by-side and domain-swapped dimers can form with relative ease. However, when lysozyme crowders are introduced, the simulations reveal a destabilization effect, particularly for the side-by-side dimer. This is evidenced by a state where a lysozyme molecule is intercalated between GB1 monomers, effectively reducing the number of direct intermonomer contacts. The observed destabilization effect from PTMetaD-WTE assisted simulation underscores the role of excluded volume interactions and possible weak attractive interactions between lysozyme and GB1 in modulating protein association equilibria in crowded environments.

Klauda and Bodosa (2024)'s usage of 2-dimensional WT-MetaD revealed critical details about the binding pathways, such as distinct metastable states (B1, B2, and B3 for BTMA) and their associated interactions, including salt bridges and π-cation interactions. It identified energy barriers and key residues stabilizing ligand binding, shedding light on the differences in pathways for the smaller, more flexible BTMA and the bulkier TPP. To enhance accuracy, umbrella sampling was also used to extend the free energy landscape analysis, providing precise measurements of changes in free energy due to solvation (ΔGsolv). The free-energy landscape analysis yielded measurements of ΔGbind and ΔGcalc which closely matched experimental isothermal titration calorimetry (ITC) values, validating the computational approach.

## Use of Enhanced Sampling Techniques to Analyze Trajectories from Biomolecular Systems

### *(1) Replica-Exchange Molecular Dynamics*
The performance of an REMD algorithm is often discussed in terms of the observed probability of a successful exchange, where a very low exchange probability will increase the computational cost of exploring the energy landscape. Once adequate sampling of the energy landscape is established by a suitable exchange probability – and thus, good replica mixing – simulations are typically analyzed in terms of the convergence of the target energetics or conformational ensemble.

The convergence is typically checked by plotting the free energy surfaces as a function of a low number of structural parameters, such as monomer radius of gyration (Rg) or nematic order parameters in the assemblies. Temperature- or distance-dependent PMFs may be calculated, providing another characteristic of aggregation propensities than what would normally be available from conventional MD simulations.

Various distributions of structural metrics may also be used to examine convergence of the conformational ensemble or close-packed aggregate structure, including radial distribution functions, intermolecular coordination distances, angles, or Rg. Contact maps may be used to characterize and compare protein conformation or aggregate packing. In unconstrained studies where an atomistic reference of the target structure is available, such as crystallographic coordinates of a bound protein, the simulated similarity to the target structure may be calculated in terms of root mean square displacement (RMSD) of the simulated atoms, observed Hydrogen-bonding, or secondary structure.

### *(2) Umbrella Sampling*
Popular methods to analyze umbrella sampling simulations and determine free energy are WHAM and umbrella integration (Kastner 2011). In WHAM, the distribution in each window is given a weight and a weighted average distribution is calculated. The weights are assigned so that the statistical error is minimized. Alternatively, in umbrella integration the mean force is averaged throughout the windows.

### *(3) Metadynamics*
Application of metadynamics procedures rely heavily on its balance between precision of fuller free energy profiles and lower time cost compared to other enhanced sampling methods, especially when dealing with systems of complex degrees of freedoms. From its inception, the authors (Piana and Laio 2007) highlighted BE-MetaD's efficiency in sampling the folded state of

Trp-cage test case, in a few nanoseconds and provided a detailed reconstruction of the free energy surface in a fraction of the time compared to REMD. The authors conclude that BE-MetaD significantly improves the exploration of complex free energy landscapes with reduced computational cost, demonstrating its potential in biologically relevant applications.

Such advantages are further accentuated in the popularity of WT-MetaD, where customization of multiple CVs further allows for analysis of conformational changes and kinetic coordinates. Recent studies explore the method's benefits in monitoring and constructing assembly pathways.

Yuan et al. (2020) explored the dissociation pathways in detail by constructing free-energy profiles and landscapes based on two key CVs: coordination number and distance between ligand and receptor. The simulations also highlighted significant differences in the dissociation energy barriers, with tafamidis having a lower barrier to dissociation, indicating a greater tendency to dissociate compared to AG10 despite its stronger binding affinity.

In Ferranari et al. (2023), the complex energy landscape associated with different coordination modes of $Zn^{2+}$ to the phosphate groups was difficult to characterize due to the high energy barriers between different configurations. WT-MetaD played a crucial role in providing an accurate structural and energetic description of $Zn^{2+}$, by selecting CVs for the coordination number of $Zn^{2+}$ to phosphate oxygens. For larger systems, Sakai (2018) employed the WT-MetaD method to explore free energy landscapes and define the interaction pathways of the ligands as they enter the M2 channel pore.

Wang et al. (2024) demonstrated a key strength of this approach to be the careful selection of CVs, which acted as RCs to represent critical interactions within the signaling complex. These included the coordination between TLR3 and dsRNA (RC1), intradimer coordination within TLR3 dimers (RC2), and interdimer coordination between dimers (RC3). These variables captured the essential configurational changes during the assembly and disassembly processes, allowing precise monitoring of the system. The study also showcased the versatility and power of metadynamics in tackling complex biological problems involving large biomolecular systems, including other pH-sensitive protein assemblies.

With increase in computing capacities, incorporation of multiple enhanced sampling methods, especially metadynamics method, has become the prospective approach. These methods target better scaling and holistic thermodynamics and kinetic representation of specific systems of interest.

Bonomi et al. (2013) combined the PT-MetaD scheme with the well-tempered ensemble (WTE) approach to mitigate the high computational cost of simulating protein binding with explicit solvents. The simulation revealed a complex free-energy landscape with a global minimum indicating favorable monomer–monomer interaction. The simulation also allowed the calculation of the binding free energy of the dimer, which agreed with experimental estimates, albeit with a higher simulated affinity due to differences in the definitions of the native state. Klauda et al. (2024) employed a 2-dimensional WT-MetaD protocol, which allowed them to investigate both the spatial and orientational degrees of freedom of the ligands BTMA and TPP. Using eight walkers to significantly enhance the sampling efficiency, they overcame challenges such as ligand trapping in specific orientations or states.

*(4) ML Techniques*

The application of artificial Intelligence (AI) methods targeted particularly towards self-assembly is limited in the scope of directly handling a multitude of kinetic pathways of self-assembly. However, there has been considerable progress in finding the right order parameter (Sidky, Chen et al. 2020) that can be used to accelerate MD simulations of self-assembly in tandem with enhanced sampling techniques.

GraphVAMPnets by Liu et al (Liu, Xue et al. 2023) has been used to build a KNM to study the kinetics of the self-assembly of hexaphenylsilole (HPS) in dimethylsulfoxide (DMSO) solvent (Liu, Kalin et al. 2024). HPS shows aggregation-induced emission (AIE), a phenomenon observed in molecules that exhibit enhanced fluorescence emission when they aggregate in the solution or the solid state. For each metastable state uncovered through the kinetic model, the AIE fluorescence intensity was estimated using QM/MM calculation on representative conformations. The overall intensity was determined by averaging across different states, weighted by population each time and was found to match the experimentally observed fluorescence intensity. Additionally, HPS aggregation through the KNM affirmed classical nucleation growth theory in which the clusters spontaneously grow beyond a certain critical size as shown by the steady relative fluorescence intensity values after some time for different initial concentrations.

GraphVAMPnet was also initially applied to the aggregation of two hydrophobic molecules (pdb id: 9d9f) in water and the self-assembly of patchy particles (Liu, Xue et al. 2023). The learned CVs of the hydrophobic molecules showed distinct binding states on the free energy surface. The states described proper orientational correlation and monotonic decrease in solvent accessible surface area towards stable valleys as the free energy decreased, indicating hydrophobic collapse of the molecules. For self-assembly of patch particles, the KNM lumped all microstates into five major macrostates when projected onto the learned two dimensional CV space corresponding to different ring-like structures that eventually formed stable 12 pentagonal rings.

Ferguson et al. (Long and Ferguson 2014) used dMap to understand the self-assembly of anisotropic patchy colloidal particles that favor forming tetrahedral and icosahedral clusters. Brownian dynamics of patchy colloidal particles along dMAp coordinates revealed distinct mechanisms of the growth of icosahedral clusters: one from the sequential monomeric addition and another from the budding of the disordered liquid structure. Whereas tetrahedral clusters are assembled through two other distinct pathways: either through the chains of stacked interlocking dimers and trimers formation or through the stacked interlocking trigonal planar trimers formation.

Herringer et al (Herringer, Dasetty et al. 2023) used permutational invariant network for enhanced sampling (PINES) that integrates permutationally invariant vector featurization with Molecular Enhanced Sampling using Autoencoders (MESA) (Chen and Ferguson 2018) to discover symmetry-adapted slow CVs. The network design employs a maximum mean discrepancy-Wasserstein autoencoder (MMD-WAE) (Zhao, Song et al. 2017) with the loss function consisting of a mean-squared error (MSE) term to enable reconstruction and a maximum mean discrepancy term to regularize latent space. The method was applied to look into the assembly/disassembly of a 13 atom Argon cluster as well as other systems, along with PB-MetaD enhanced sampling using learned CVs.

Bonati et al. (Bonati, Piccini et al. 2021) used Deep-TICA with OPES (Invernizzi and Parrinello 2020) simulation to study silicon crystallization which is a first-order phase transition hindered by a large free energy barrier. Similarly, this method can also be applied to crystallization-driven self-assembly of block copolymers through a first-order phase transition. Qiu et al (Qiu, O'Connor et al. 2023) used the variational autoencoder-based flux approach to efficiently lump parallel kinetic pathways elucidated from the Markov state model (MSM) and Transition Path Theory (TPT) into distinct metastable path channels. These channels helped comprehend the molecular mechanism underlying the conformational changes in the system. Latent space path clustering (LPC) was employed to understand the aggregation of two hydrophobic molecules (pdb id: 9d9f) in water.

Evolutionary reinforcement learning was used by Whitelam (Whitelam and Tamblyn 2020) to optimally control parameters such as temperature and chemical potential to guide the self-assembly process. A disk model was employed to simulate real-world systems, like colloidal self-assembly, effectively capturing competition between polymorph formation and far-from-equilibrium dynamics.

## Section 5 - Conclusions and Outlook

Biomolecular assembly plays a critical role in physiology and the technology supporting health care. Despite its ubiquity in nature and laboratory, identifying thermodynamically stable self-assembled structures within a global minima or structures of rare molecular conformation is difficult through computational methods. This is due to a variety of reasons pertaining to limitations in time scale and energy barriers. In addition, determining free energy differences between these conformational states to identify driving forces is just as difficult. Since traditional computational methods cannot parse these biomolecular states and energy differences, non-traditional or "enhanced" methods have to be applied. These methods consist of REMD, metadynamics, umbrella sampling, and ML/AI.

Replica exchange is a robust and well-tested technique to enhance sampling of macromolecular conformations and intermolecular interactions in biomolecular systems. The simplicity of the core idea of RE has allowed for a fascinating variety of implementations, allowing researchers to judiciously modulate system energy by harnessing a variety of model features, especially in MD models. Such experiments have led to a number of findings relevant to supramolecular self-assembly in biomolecules, including the identification of assembly pathways for amyloid peptide aggregates and protein quaternary structures. Umbrella sampling is a versatile method that aims to determine free energy differences and determine possible driving forces underlying self-assembly. It can be applied stand alone or combined with other techniques such as REMD. Notable papers applied umbrella sampling to peptides and polymer systems to determine the mechanisms underlying self-assembly and the contributions from each type of force. Aside from incorporating other enhanced sampling methods and quantum mechanics to further expand its applicability and improve continuity across resolutions, recent innovations in metadynamics emphasize collaboration with both intradisciplinary and interdisciplinary topics. One of the most discussed is the area of pattern recognition. Sampath et al. (2020) suggested developments in modeling techniques incorporating metadynamics that have advanced our understanding of the molecular recognition patterns in systems that are dictated by self-assembly, exemplifying systems such as biomineralizing peptides, hierarchical peptoid assemblies, and large protein assemblies. While also pointing out prospects with additions in statistical learning and AI, the integration of ML to optimize CVs (Bonati et al. 2021) and hybrid approaches combining metadynamics with experimental data represent promising advancements.

While this review article discusses some commonly used enhanced sampling methods, there are several other methods, for example, adaptive bias implementations such as variationally enhanced sampling and Wang-Landau (Henin et. al. 2022). Similarly, complementary to adaptive bias potential methods, there are also adaptive bias force methods which seek to add a force component to level the free energy barriers. Other implementations include adaptive sampling and simulated annealing (Henin et. al. 2022; Bernardi et. a. 2015).

It can even be argued that coarse-graining is an enhanced sampling method for resolving the self-assembly process (Spiwok et. al. 2015). Coarse-graining a set of atoms into a bead to act as a "pseudo atom" reduces the degrees of freedom, smoothens the potential energy landscape, and increases the timescale of the simulation (Aydin et. a. 2014; Aydin and Dutt 2014; Spiwok et. al. 2015; Aydin and Dutt 2016; Mushnoori et. al. 2018; Mushnoori et. al. 2020; Banerjee et. al. 2021; Banerjee and Dutt 2023; Mushnoori et. al. 2023; Banerjee et. al. 2024). This approach allows for sampling of new conformations although not at atomistic resolution. To address this, there are, however, backmapping and clustering methods to assess the similarity between coarse-grained and atomistic self-assembled structures (Banerjee et. al. 2021; Banerjee and Dutt 2023; Hooten et. al. 2023). A recent study performed bottom-up coarse graining of tri-phenylalanine to extend the simulation time from nanoseconds to microseconds and assessed the final coarse-grained structure using backmapping (Hooten et. al. 2023).

Finally, even though enhanced sampling greatly aids the understanding of self-assembled biomolecular materials, self-assembly is not yet solved. There are still limitations in time scale and length scale that need to be addressed. Improvements in computer architecture and compute power can slowly address these concerns over the years (Spiwok et. al. 2015). Additionally, there has been a push to intermix and combine a variety of enhanced sampling techniques to form better combined sampling methods. There has also been a push to combine ML and AI in conjunction with a variety of enhanced sampling techniques to discover CVs that optimally span free energy surface submanifold, accounting for long-lived, dominant metastable states of kinetic pathways (Sidky, Chen et al. 2020) (Mehdi, Smith et al. 2024). Furthermore, there are several ongoing efforts to decouple the enhanced sampling techniques from the respective software and develop efficient enhanced sampling implementations that are universal so that they can be used on data from community-based molecular simulation packages.

Given the prevalence of the process of self-assembly in a wide range of disciplines, there continues to be numerous efforts to develop unique methods and iterate upon them to advance these disciplines. This review aims to provide an introduction to some common enhanced sampling methods which are applied to develop a fundamental understanding of the process of self-assembly, and provides an outlook on how some of these techniques can potentially evolve.